# A Work Zone Simulation Model for Travel Time Prediction in a Connected Vehicle Environment

Xuejin Wen*

Ohio State University
* wen036@gmail.com

**Abstract**
A work zone bottleneck in a roadway network can cause traffic delays, emissions and safety issues. Accurate measurement and prediction of work zone travel time can help travelers make better routing decisions and therefore mitigate its impact. Historically, data used for travel time analyses comes from fixed loop detectors, which are expensive to install and maintain. With connected vehicle technology, such as Vehicle-to-Infrastructure, portable roadside unit (RSU) can be located in and around a work zone segment to communicate with the vehicles and collect traffic data. A PARAMICS simulation model for a prototypical freeway work zone in a connected vehicle environment was built to test this idea using traffic demand data from NY State Route 104. For the simulation, twelve RSUs were placed along the work zone segment and sixteen variables were extracted from the simulation results to explore travel time estimation and prediction. For the travel time analysis, four types of models were constructed, including linear regression, multivariate adaptive regression splines (MARS), stepwise regression and elastic net. The results show that the modeling approaches under consideration have similar performance in terms of the Root of Mean Square Error (RMSE), which provides an opportunity for model selection based on additional factors including the number and locations of the RSUs according to the significant variables identified in the various models.  Among the four approaches, the stepwise regression model only needs variables from two RSUs: one placed sufficiently upstream of the work zone and one at the end of the work zone.

**KEYWORDS**:
Connected Vehicle, Road Side Unit, Work Zone Travel Time, PARAMICS

**Introduction**
  The ability to provide accurate traffic state predictions in real-time has long been identified as a critical need for Intelligent Transportation Systems (ITS), especially under abnormal conditions such as traffic accidents *(1)* and inclement weather *(2)*. Work zone (WZ) is a main issue that may lead to abnormal traffic congestions. It typically requires one or two lanes of the freeway segment closed, resulting in temporary capacity reduction. Based on the report of Federal Highway Authority (FHWA), just in 2014 nearly 24 percent of non-recurring freeway delay, equivalent to about 888 million hours, is typically attributed to work zones *(3)*. Work zone may also lead to safety issues. In 2015 New York State DOT (NYSDOT) reported 646 crashes in capital construction work zones on New York roads and bridges, resulting in six motorist fatalities and 147 injuries to motorists, contractor employees and NYSDOT staffs *(4)*.

  A few studies have focused on providing current or predicted travel times through WZs in the freeway network. For example, the variable message signs (VMS) can display the travel time and distance to the end of the work zone to motorists *(5)*. Lee and Kim (2006) designed and implemented an Automated Work Zone Information System (AWIS) to provide road users with real-time delay information during peak hours so that the traffic can be diverted to detour routes *(6)*. Ghosh-Dastidar and Adeli (2006) built a neural network model using simulated data to predict the delay and queue length at freeway work zones *(7)*. Bae et al. (2017) also applied a neural network model to predict the traffic volume and percentages of trucks at the upstream of the work zone, then they applied a BPR function to calculate the work zone travel time based on the ratio of volume and capacity *(8)*.

  The estimation and prediction of travel times requires measurement of traffic states. Traditionally the traffic detectors are installed on freeways to record traffic characteristics such as flow, speed and occupancy, which are used as the data source for travel time prediction *(9)*. For example, California Freeway Performance Measurement System (PeMS) has installed 43,426 detectors on freeways with a total distance of 30,599.4 miles by 2017. However, based on the latest statistics, 37.1% of PeMS detectors have been out of function *(11)*. The installation and maintenance costs of these huge amounts of fixed detectors are therefore very high. On the other hand, if the deployment is not dense, there may not be enough detectors around the work zone segment to collect the necessary information.

  Connected vehicle is an emerging technology recently. It allows each vehicle to be fully connected to the other vehicles and to the infrastructures like traffic light and Variable Message Sign (VMS). The U.S. Department of Transportation has made the development of vehicle-to-vehicle (V2V) and vehicle-to-infrastructure (V2I) communication a major research priority to improve the safety and efficiency of the transportation system. Connected vehicles can store their vehicle activities such as speed, position, turn signal, brake status and so on with a frequency in On-board Units (OBUs). At the same time, there are roadside units (RSUs) installed at intersection, interchanges and other locations to provide communication interface to the vehicles. When a connected vehicle passes by the coverage area of a RSU, the vehicle activities data in OBU is transmitted to RSU and the OBU buffer will be cleared *(9)*.

  In this study, we propose to utilize portable roadside unit (RSU) for freeway work zone travel time prediction in a V2I environment. Comparing with the fixed detectors, the locations and numbers of the reusable and portable RSUs can be adjusted based on work zone types, traffic demand and the connected vehicle penetration rates. Simulation is an important approach for this kind of experiments with advantages like safe, convenient and almost no cost *(10)*. Here a real-world work zone simulation model in connected vehicle environment is built using PARAMICS®, microsimulation software by QuadstoneParamics. The work zone travel time prediction models are also built to identify candidate locations of RSUs and verify the connected vehicle techniques can improve the travel time prediction accuracy.

  The rest of the paper is organized as follows. The next section will introduce the PARAMICS work zone simulation model for NY State Route 104. Following that, the simulation results are processed and relevant variables are extracted. The work zone travel time prediction models are then built based on these variables. Conclusions and recommendations are given at last.

**Work Zone Simulation Model**
  The configuration of the work zone simulation model is observed from NY State Route 104. As shown in Figure 1, it is a piece of two-lane freeway with no ramps and the traffic are from left to the right. The length of the one-lane-closed segment is 2 miles. A warning arrow panel and a VMS sign are located at 0.1 miles and 0.7 miles from the work zone beginning separately.

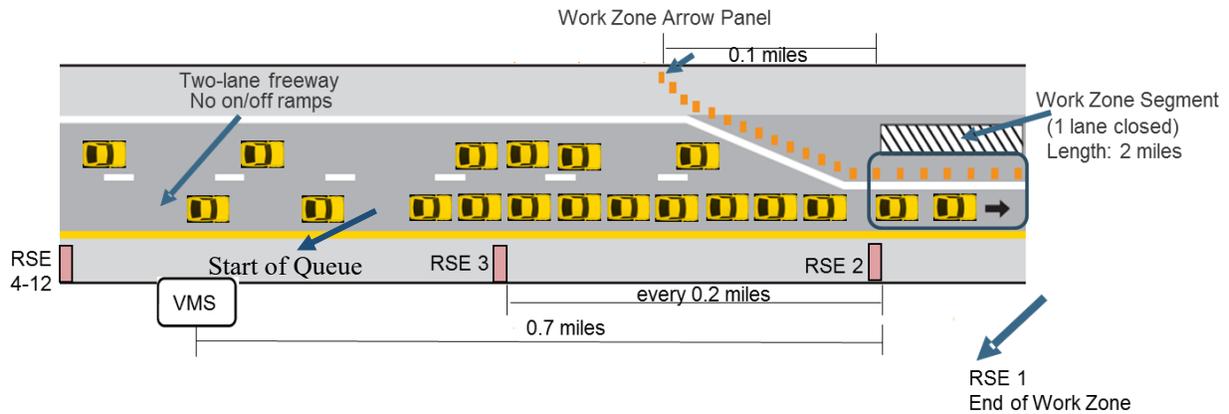

**Figure 1. Layout of WZ from NY-104**

After that, the simulation model was built using PARAMICS *(12)*. Twelve portable RSUs were introduced in this model (Figure 1). RSU 1 is at the end of the work zone, the rest RSUs 2-12 are distributed every 0.2 miles from the beginning of work zone to 2 miles at the upstream of work zone. Including a dense set of RSUs in the simulation enables us to explore the impact of RSU location on travel time prediction. As will be discussed, through the development of work zone travel time prediction models, the significant variables as well as the RSUs that provide those information can be identified, and therefore the number and locations of the RSUs also become obvious.

The hourly traffic volume data in 2014 for this freeway segment are acquired from NY Department of Transportation *(13)*. As shown in Figure 2, for each month, we took the average by hour and got the typical traffic volumes. Then for the 24-hour traffic volumes of each month, we conducted 10 runs of simulations with different random seeds so that stochastic behaviour is considered. In this study, the market penetration rate of connected vehicles was set to 100%. The PARAMICS API was implemented in C++ to simulate the V2I applications and collect the data.

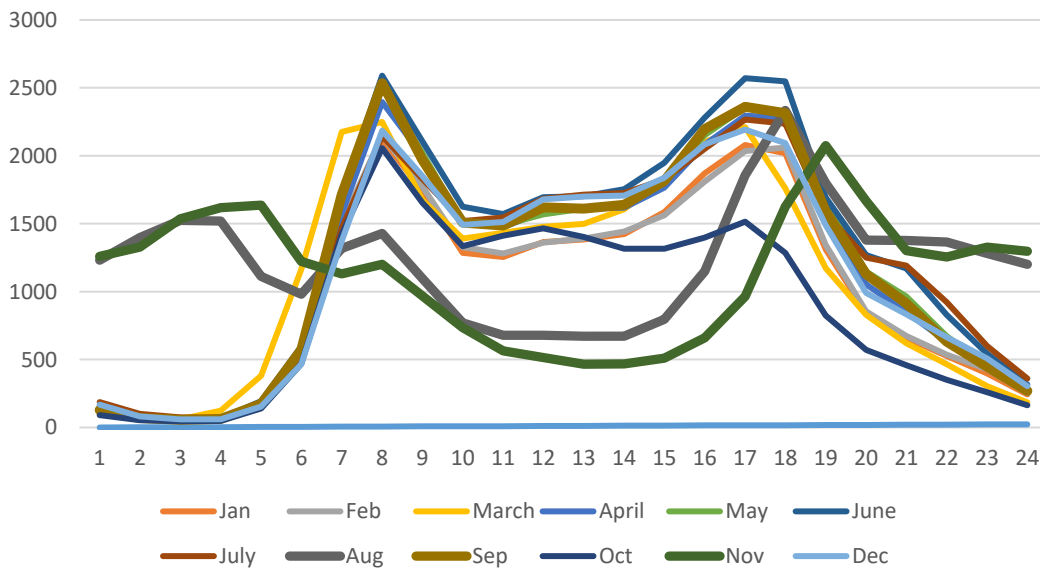

**Figure 2. Hourly Traffic Volumes Averaged by Month**

**Simulation Data Processing**

In this connected vehicle simulation model, the RSU has the function to record the ID and timestamp when a vehicle passes by it. The vehicle will also transmit its previous trajectory data to the RSU. Finally, we extract the following variables from the simulation data. Note that all these variables are aggregated by five minutes.

a) $TT_{i,1}, i = 2,3,...,12$. Based on the vehicle ID and time stamp, we calculate the travel time from each of the upstream RSU to the one at the end of the work zone. For example, "$TT_{2,1}$" is the travel time from RSU 2 to RSU 1.

b) *UpstreamFlow*. The traffic volume at the upstream RSU 12.

c) *DownstreamFlow*. The traffic volume at the downstream RSU 1.

d) *WorkZoneEndAcc*. The acceleration rate at the end of work zone, which is based on the vehicle's trajectory collected by RSU 1.

e) *Qlength*. The length of the queue at the upstream of the work zone. In this study, a queue is defined to satisfy three criteria: the speed is less than 10 miles/hour, the inter-vehicle distance is less than 0.01 miles and the number of vehicles is at least 5. The queue information can be extracted from the trajectory data stored in RSU 1.

f) *StartQueue*. The location of the start of the queue as shown in Figure 1.

g) $TT_{wz}$. The work zone travel time. Note that for most of the previous studies, they measured the work zone travel time from a fixed location at the upstream to the end of the work zone *(5) (7) (8)*. Although this simplifies the problem, however, it is not accurate because the length and location of the upstream queues are changing. The approach we measure $TT_{wz}$ is as follows. If there was a queue, it is the time from the leftmost vehicle in the queue to the end of work zone. If there was no queue, it is the travel time through the exact work zone.

With the five-minute aggregation, for 24-hour traffic volumes of each month, we can get 2,880 observations of variables a) - g) by running the simulation model 10 times. In total, 28,800 records from the first 10 months (January to October) are used for the training of the travel time prediction model, and the rest 5,760 records from November to December are used for the testing of the model performance.

**Work Zone Travel Time Prediction and RSU Location**

Let $TT_{wz}(t)$ denote the travel time through a work zone at time step t and *Lahead* be the "look ahead" time. In this study, *Lahead* is set at 5 time steps, which corresponds to predicting travel time through the work zone 25 minutes into the future (data is aggregated every 5 minutes, so 5 time steps equates to 25 minutes). To predict the future travel time, we use an autoregressive with exogenous input model structure as follows:

$$TT_{wz}(t + Lahead) = A(q)TT_{wz}(t) + \sum_{i=1}^{n_u} B_i(q)u_i(t); \qquad (1)$$

where $A(q) = \sum_{j=1}^{n_a} a_j q^{-j}$, $B_i(q) = \sum_{j=1}^{n_{b_i}} b_j q^{-j}$;

$u_i$ refers to one of 16 variables (11 travel time variables $TT_{i,1}, i = 2,3,...,12$, *UpstreamFlow, DownstreamFlow*; *WorkZoneEndAcc, Qlength* and *StartQueue*), $n_u = 16$;

$A(q)$ and $B_i(q)$ are the backshift operators, which operates on an element of a time series to produce the previous element;

$n_a$ and $n_{b_i}$ are the backshift step lengths;

The coefficients $a_j$ and $b_j$ are derived using the algorithm that predicts travel times in the work zone. In this study, we applied the following four regression techniques: simple linear regression model; multivariate adaptive regression splines (MARS), with up to 2nd order nonlinear interactions in the input variables *(14)*; a stepwise model with risk inflation criterion *(15)*; and an elastic net model *(16)*. All the experiments are conducted using statistical software R.

Table 1 summarizes the significant variables selected by the four models. The root mean squared error (RMSE) between the real and predicted work zone travel times is also calculated for each model.

**Table 1. Significant Variables and RMSEs of the Four Models**
**(* means that variable is significant for that model)**

| | | Models | | | |
|---|---|---|---|---|---|
| | | Linear Regression | MARS | Stepwise | Elastic Net |
| 16 Variables | $TT_{2,1}$ | | | | |
| | $TT_{3,1}$ | | | | |
| | $TT_{4,1}$ | | | | * |
| | $TT_{5,1}$ | | | | * |
| | $TT_{6,1}$ | | | | * |
| | $TT_{7,1}$ | | | | * |
| | $TT_{8,1}$ | | * | | * |
| | $TT_{9,1}$ | | | | * |
| | $TT_{10,1}$ | * | * | | * |
| | $TT_{11,1}$ | | | | * |
| | $TT_{12,1}$ | * | * | * | * |
| | *UpstreamFlow* | * | * | * | |
| | *DownstreamFlow* | | | | |
| | *WorkZoneEndAcc* | | * | | |
| | *Qlength* | | | | * |
| | *StartQueue* | * | * | | * |
| RMSE (seconds) | | 31.46 | 28.37 | 34.64 | 30.82 |

As can be seen in Table 1, first for the variable selection, the linear regression model selects four significant variables, which are collected by three RSUs 1, 10 and 12; the MARS model identifies six significant variables from RSU 1, 8, 10, 12; the Elastic Net model is the most complicate one with 11 significant variables from all the other RSUs except 2 and 3; and the Stepwise model is the simplest one with only two significant variables from RSU 1 and 12. One interesting observation is that the variables from the RSU 10 and RSU 12 are the most significant ones, e.g. $TT_{10,1}$ is selected three times and $TT_{12,1}$ are included by all the four models. This tells that the location and length of the queues caused by the work zone are probably around RSU 10 and RSU 12.

Second, for the travel time prediction performance, MARS model performs the best with a RMSE of 28.37 seconds. It's a little surprising that the Elastic Net model has a little worse RMSE of 30.82 seconds with the requirement of so many variables to build the model. The linear regression model ranks the third, its RMSE is 31.46 seconds. Finally the RMSE of the simplest Stepwise model with two variables is around 34.64 seconds.

Third, we can also make the portable RSU location plans based on Table 1. The performance of Elastic Net model is not the best and it needs 10 RSUs, which is unrealistic. Therefore this plan can be discarded. For the other three models, there is a trade-off between the algorithm accuracy and the RSU resources. The MARS model has the smallest RMSE but it needs variables from 4 RSUs. Its performance is followed by the linear regression model which requires 3 RSUs. The RMSE of the stepwise model is only about 6 and 3 seconds higher than MARS and linear regression model, respectively. And it just requires two RSUs: RSU 12 and RSU 1, which, from a practical viewpoint might represent a reasonable trade-off between algorithm performance and sensing requirements.

We are interested to further check the performance of the Stepwise Model. Figure 3 shows the real work zone travel time (red) and the predicted one (blue) for the testing set using this Stepwise model. Figure 4 shows the histogram of the error between the real work zone travel time and the predictions.

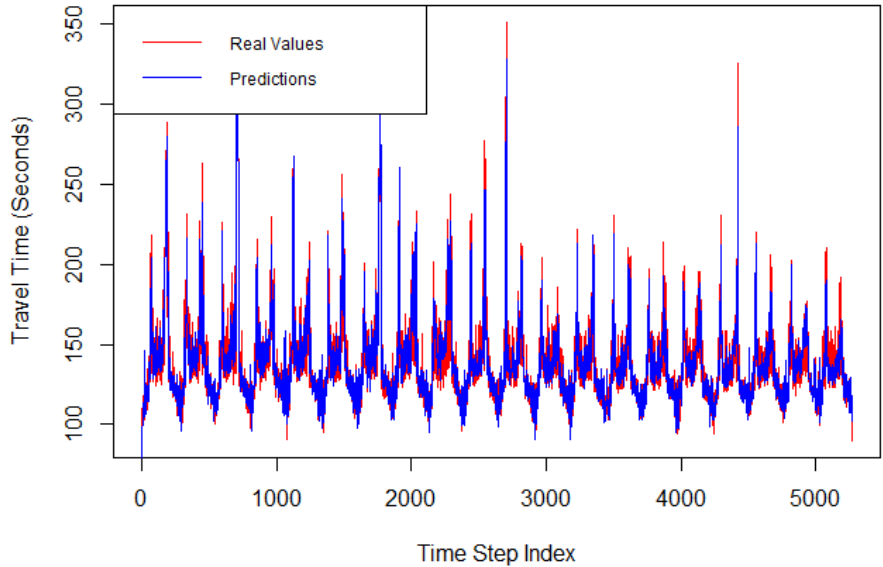

**Figure 3. Real Travel Times and Stepwise Model Predictions for the Testing Dataset**

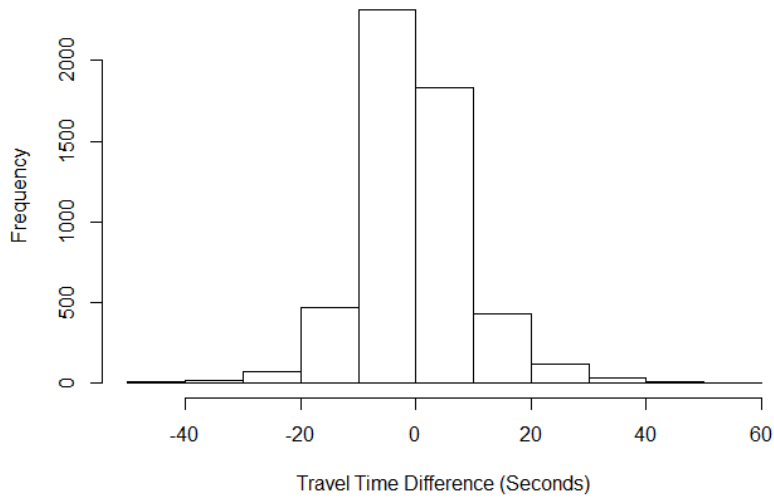

**Figure 4. Histogram of Travel Time Difference between Real Values and Predictions**

Figure 3 shows that in general the Stepwise model can capture the trend of work zone travel times very well. However, it looks like that the predictions from Stepwise model (blue line) are lower than the real work zone travel times (red line) when the ground truth values are high. Figure 4 shows the distribution of the difference between the real travel times and the predictions. We can see that the frequency with the absolute difference less than 10 seconds is very high. Most of the time, the difference is within [-20, 20] interval. This once again shows the Stepwise model has a reasonable performance.

**Conclusion**

This paper focuses on travel time prediction for freeway work zones in a connected vehicle environment. A work zone segment motivated by NY State 104 was simulated using PARAMICS. The simulation model considered a Vehicle-to-Infrastructure setting by using portable RSUs to communicate with connected vehicles and collect traffic data. Compared with loop detectors, the portable RSUs can save installation and maintenance cost, and their number and locations can be adjusted. To determine candidate RSU locations, twelve RSUs were put along the work zone segment in the simulation model and sixteen traffic variables were extracted from the V2I data. For travel time prediction and variable selection, four different modelling approaches were investigated, including linear regression, MARS, stepwise and elastic net. The results showed that the models have similar RSME performance, ranging from 28 seconds to 35 seconds. Through the comparison of the identified significant variables, a suggested number and locations of the portable RSUs were also determined. We found that the stepwise model provides a practical sensing plan in that only two RSUs are required: one sufficiently upstream from the work zone and one at the end of work zone.

For our future study, more experiments are necessary. For example, we can explore more scenarios with different connected vehicle penetration rates and work zone types. We can also adjust the RSU location plans dynamically based on various traffic demands by peak-hour and non-peak-hour, weekday and weekend, inclement and clear weather and so on.